\documentclass[a4paper,11pt]{article}
\usepackage[T1]{fontenc}
\usepackage[utf8]{inputenc}
\usepackage{lmodern}
\usepackage{graphicx}
\usepackage{epstopdf} 
\usepackage{color}
\title{A path to confine gluons and fermions through complex gauge theory}

\author{R. L. P. G. Amaral$^{a}$\footnote{email: rubensamaral@id.uff.br} ,
	V. E. R. Lemes$^{b}$\footnote{email: verlemes@gmail.com},
	O. S. Ventura$^{c}$\footnote{email: ozemar.ventura@cefet-rj.br}, L.C.Q.Vilar $^{b}$\footnote{email: lcqvilar@gmail.com}  \\
	\small \em $^a$Instituto de F\'{\i}sica, Universidade Federal do Fluminense\\
	\small \em Av. Litor\^anea S/N, Boa Viagem, Niter\'oi-RJ CEP. 24210-340,
	Brazil\\
	\small \em $^b$Instituto de F\'\i sica, Universidade do Estado do Rio de
	Janeiro,\\
	\small \em Rua S\~{a}o Francisco Xavier 524, Maracan\~{a}, Rio de Janeiro - RJ,
	20550-013, Brazil\\
	\small \em $^c$Departamento de F\'\i sica, Centro Federal de Educa\c{c}\~ao Tecnol\'ogica do Rio de
	Janeiro\\
	\small\em Av.Maracan\~a 249, 20271-110, Rio de Janeiro - RJ, Brazil}

\begin{document}  

\maketitle

\begin{abstract}

We introduced here the study of a QCD based on a complex group. Our aim is to show that a gauge theory with a complex symmetry develops some of the features required for the description of a confined phase. This theory leads to gluons with propagations characteristic of i-particles, fundamental constituents of propagators in the Gribov scenario, together with a gluon propagating with the mode predicted by 't Hooft for the quark confinement. In this way, we present in this same theory a possible gluon condensate, together with a confining interquark potential after the coupling to fermions. This is a novelty, as, up to now, all theories of the GZ kind do not generate quark confinement.

\end{abstract}

\section{Introduction}

Along the construction carried out by Zwanziger of a field theory for the gluon dynamics encompassing Gribov's ideas \cite{Gribov} for the confinement (GZ theory) \cite{Zw89}, several principles emerged as cornerstones for further developments (a review on this topic with an extensive list of references is found in  \cite{Zw12}). Among them, we highlight the relation between confinement and the loss of positivity, which in itself explains the fact that the fundamental fields propagators of the theory do not directly define a physical spectrum. We must also mention the original idea of creating a theoretical environment for confinement from the addition of purely trivial BRST elements to Yang-Mills lagrangian, a topological sector. The main goal was then to perceive confinement as an inevitable outcome of pure Yang-Mills theory. Good agreements with the lattice approach to the gluon confinement should also be remarked \cite{Dud08}.

However, in order to create dynamics from this topological structure, a process of fixation of BRST sources is embedded in Zwanziger's scheme \cite{Zw93}. This ultimately leads to a soft breaking of the BRST invariance \cite{So09}. In fact, the instability of the BRST symmetry in the non-perturbative regime was understood a long time ago \cite{Fj83}.  Although the renormalizability of the GZ theory has been checked in the Landau gauge \cite{Ted11}, it soon became clear that this process implied a gauge dependent construction \cite{Lav11, Lav12}. Each gauge fixing would require a different horizon function, which would intrinsically carry this gauge dependence \cite{Lav15}. 
The acknowledgement of this inconsistency led to the further split of the development of the GZ program into two main lines. One searched the idea of immersing this original structure into a more complex field architecture, allowing for the extension of the GZ theory with an exact BRST symmetry. This was based in a previous observation of a non local exact BRST symmetry of GZ in the Landau gauge \cite{Sor09, Kon09}. The theory was adapted, firstly to develop the same non local symmetry in a general linear covariant gauge \cite{Cap15}, and then in order to localize the BRST symmetry into an infinite polynomial form \cite{Cap16}. This process is based on the introduction of a Stuckelberg field in the same way as it was once proposed as an alternative for the non abelian Higgs mass picture of Yang-Mills theory \cite{Del86}. At that time, this proposal was shown to be inconsistent \cite{Kubo87, Kos87}, as a conflict between renormalizability and unitarity was proven inevitable \cite{Del88}, a situation which remains up to now (see for example the review \cite{Rue04}). Another issue of this new version of GZ theory is that the action itself becomes an infinite series, imposing the interpretation of a weak coupling expansion above a non perturbative vacuum, as it does not converge for a strong coupling phase \cite{Cap18}.

Simultaneously, another line of research grew from another even older observation: there is a way to interpret the full GZ theory as a spontaneous symmetry breaking of BRST starting from pure Yang-Mills plus the trivial topological sector \cite{Mag94}. The mechanism itself generating this phase transition is not described, but the broken phase is characterized by the existence of a non null vacuum expectation value of a BRST trivial element. Some problems with the derivation of the GZ theory from this symmetry breaking point of view were then pointed out, mainly from the fact that the whole approach depended explicitly on the space time coordinate. Its inconsistency was argued in \cite{Dud08}. Since then, this criticism was circumvented by different proposals. Some of the authors proposed in \cite{Vil11} a spontaneous symmetry breaking mechanism, without any space time dependence, based on Fujikawa's work \cite{Fj83}, generating a refined Gribov propagator with an abelian dominance, an effect predicted by the lattice \cite{Cuc11}. Alternatively, the implementation of a finite volume quantization was carried out in \cite{Zw12, Sc15}, in a way to avoid the inconsistency of the original Maggiore-Schaden construction. And another spontaneous symmetry breaking effect, also free from the explicit space time dependence, again by the extension of the field content of the theory, was presented in \cite{Sor12}. Nevertheless, these options were only developed for the specific Landau gauge (as it was actually recognized in \cite{Sor12}), and so may suffer from the same gauge dependence problem that motivated the analysis in \cite{Lav11, Lav12} (also the existence of preferred directions on the resulting vacuum seems to be another drawback).

This state of the art briefly resumed does not show a conclusive route to overcome all the theoretical problems. Anyway, all these developments brings us the impression that GZ could in fact be describing a phase of a larger theory, when it is already confined. At the same time, it is natural to expect that a process of spontaneous symmetry breaking should play a major role, since this is the mechanism that we find in theoretical physics that can make this transition and at the same time preserve renormalizability and unitarity. Then, we feel that there is room to look for alternative views. Our search will be for a theory that after a spontaneous symmetry breaking leads to a confined phase in the sense of the GZ theory, but defined in a class of linear covariant gauges. 


On the other hand, as it is cleared stressed in \cite{Sor12}, the preservation of a BRST symmetry does not guarantee that a unitary description will be reached. One should in the end verify if the theory allows for a positive norm subspace of the confined degrees of freedom. As in a confined phase, the elementary excitations are expected not to have asymptotic states, this subspace must be searched among the two point correlators of composite operators constructed from the basic fields. In fact, it is very hard to describe correlators with these properties in the GZ theory. Several developments lead to the foundation of the important concept of i-particles, fundamental elements from which condensates satisfying positivity criteria would be possible \cite{Bau10, Cap11, Dud10, Dud11}. Propagators of i-particles would be behind the formation of Gribov propagators. However, the fact that the fundamental fields in GZ do not precisely represent i-particles led to the impossibility of defining condensates with the necessary properties to describe physical observables in the theory \cite{Zw89}. A new idea to overcome this obstacle was the further development of the replica model  \cite{Sor10}. Created in order to associate fundamental fields directly to i-particles, it showed once more the loss of positivity as an inevitable outcome in this confinement scenario. But another ingredient appeared: a complex gauge field is hidden in the replica model. 

Complex gauge fields show up once in a while in the physics literature. Wu and Yang used complex gauge fields in the study of instantons and duality theory, and raised the question of "a possible physical meaning to a gauge theory where the energy is necessarily not positive definite" \cite{Wu76}. There is also the famous result that three dimensional general relativity is related to complex Chern-Simons theory, where this discussion is avoided because the Hamiltonian in this theory is zero, and otherwise, as underlined by Witten, "these gauge groups would be forbidden in ordinary Yang-Mills theory" \cite{Wit91, Wit10}. Another example is the description of the field content of $N=4$ supersymmetric Yang-Mills field theory from the complexification of the $N=2$ theory \cite{Bau09}. This example is instructive in the sense that in order to restore positivity, part of the complex gauge field is constrained in such a way that it becomes an ordinary vectorial matter field. In fact, what all these authors point out is that a complex gauge field theory symmetric under a complex group leads to the loss of positivity. Then, if we take complex gauge field theory as an appropriate environment to the description of i-particles, we need to deepen the analysis. As already mentioned, we will be led to the conclusion that fundamental fields are not associated to asymptotic particle states. This can be seen as a precept to the confinement, but the main issue is that we must recover the physical spectrum of excitations of the theory, or else such theory will remain physically meaningless. A possible path opens if we assure the possibility of defining condensates from such fields. In fact, the concept of i-particles is born inside the GZ theory as a building block for condensates. These would be formed from vertices joining simultaneously pairs of i-particles and anti-i-particles. Objects built in this way would propagate respecting K\"all\'{e}n-Lehmann spectral representation, and then would define candidates for observable states. And according to what is derived in the replica model, this combination in pairs of i-particles and anti-i-particles is essential for the success of the construction.

From this point on, we derived two main possible constructions of a complex gauge field theory. One is found on a standard complex symmetry for all the fields that take part on the theory. It presents some candidates for a gluon condensate. This theory will be fully exhibited in a forthcoming work.  It begins with the usual transformation of a complex gauge field $ \mathcal{A}_{\mu }$ in the adjoint representation

\begin{equation}
\mathcal{A}_{\mu } \longrightarrow  \mathcal{A}_{\mu }^{'} = G^{-1} \mathcal{A}_{\mu } G + {i \over g} G^{-1}(\partial_{\mu } G) \label{agaugetransf}
\end{equation}
and, in the case of a complex group, we have the possibility of defining a conjugated field  $\bar{\mathcal{A}}_{\mu }$ transforming as

\begin{equation}
\bar \mathcal{A}_{\mu } \longrightarrow  \bar \mathcal{A}_{\mu }^{'} = G^{\dagger} \bar \mathcal{A}_{\mu } G^{\dagger -1} + {i \over g} G^{\dagger} (\partial_{\mu } G^{\dagger -1}) \label{abargaugetransf}
\end{equation}
since in a complex group $G^{-1} \neq G^{\dagger}$. In the next step , we can define covariant curvatures $\mathcal{F}_{\mu \nu}$ and $\bar \mathcal{F}_{\mu \nu}$ for ${\mathcal{A}}_{\mu }$ and $\bar{\mathcal{A}}_{\mu }$ respectively. Thus we immediately  see that invariant objects built exclusively from these curvatures, will be holomorphic in the sense that once ${\mathcal{A}}_{\mu }$ is found in a monomial, it will not contain $\bar{\mathcal{A}}_{\mu }$, and vice-versa. For example, we will be able to define curvature invariants as $ Tr \mathcal{F}^2$ and $ Tr {\bar \mathcal{F}}^2$. Therefore the inevitable loss of positivity. Positivity would only be ensured in the non-holomorphic element $ Tr \mathcal{F} \bar {\mathcal{F}} $, which is not invariant by the complex gauge transformations (\ref{agaugetransf}) e (\ref{abargaugetransf}). Naturally, an i-particle will be associated to the ${\mathcal{A}}_{\mu }$ field, and an anti-i-particle to $\bar{\mathcal{A}}_{\mu }$. 

We also assumed that the GZ theory should be embedded in a larger theory with a spontaneous symmetry breaking sector. The usual option is to add a scalar sector to the theory. The first idea would be to work with a complex scalar field and its conjugate. Then we can follow the holomorphism of the theory, bringing it to the scalar sector. But there is an alternative path.

This comes from the freedom that the complex group gives us. The fact that $G^{-1} \neq G^{\dagger}$ allows us an unorthodox  proposition for an adjoint inspired  scalar field transformation

\begin{equation}
\phi \longrightarrow  \phi^{'} = G^{\dagger} \phi G \label{phigaugetransf}
\end{equation}
and also
\begin{equation}
\psi \longrightarrow  \psi^{'} = G^{-1} \psi G^{\dagger -1} \, .\label{phibargaugetransf}
\end{equation}

These scalar fields, in isolation, do not form invariants. But joining them together we find $Tr \phi \psi $, which is invariant under this action of the complex group. 

Here we need to take some care to understand what is implied in the transformations (\ref{phigaugetransf}) and (\ref{phibargaugetransf}). In these it is implicit a left and a right action on the field, which indicates that the field carries a representation of the algebra, as it occurs in the traditional adjoint representation. However, as the transformations in (\ref{phigaugetransf},\ref{phibargaugetransf}) involve $ G^{\dagger}$ and not $G^{-1} $, we can conclude that for a general group these transformations are not closed inside the vectorial space of the algebra. This means that the admissibility of this pattern of transformation will depend on specific choices of the gauge group. Since  $G^{-1} \neq G^{\dagger}$, we know that unitary groups are not eligible. Nevertheless, the transformations (\ref{phigaugetransf},\ref{phibargaugetransf}) preserve hermiticity, $i.e.$, if $\phi $ and  $ \psi $ are hermitians, then so will be $\phi ' $ and  $ \psi ' $. An hermitian basis for the algebra is characteristic of the real unitary groups, but in the complex extension such basis is acceptable for the $SL(N,C)$ groups. For instance, in the case of the complex $SL(2,C)$, we can take the Pauli matrices as the generator basis, or Gell-Mann matrices for the complex $SL(3,C)$. So, let us assume that we are working with a complex $SL(N,C)$ as the gauge group. Even so, it is not still warranted that in (\ref{phigaugetransf},\ref{phibargaugetransf}) a hermitian matrix of the $sl(N,C)$ algebra will be rotated into another algebra matrix by the action of the $SL(N,C)$. The minimal cost to obtain a consistent construction is to impose that such matrices of the  $sl(N,C)$ algebra used to define the scalar fields belong to the fundamental representation. In this case, we will have a basis with  $N^2 -1$ $N \times N$ matrices, and consistency will be achieved in (\ref{phigaugetransf}) and (\ref{phibargaugetransf}) if we suppose the existence of a further $\phi^0 $ component associated with the $N \times N$ identity (and the same for $ \psi $).

This reasoning will be exposed from another perspective when we present the BRST invariance of the theory, in section II. In section III we will show the theory with asymmetric vacua taking us to i-particles and 't Hooft propagators, in the specific case of the complex $SL(3,C)$ symmetry. In section IV, we will present a candidate for the condensate that arises due to the non-holomorphism of the scalar sector of the theory, and that after a choice of the non-symmetric vacuum will present the form indicated by the replica model. Then we show how such condensate leads to a K\"all\'{e}n-Lehmann type spectrum describing a glue-ball. In section V we will couple this theory to a fermionic sector. As a major novelty of this proposal, we will show how this theory in this phase that generates the candidate for a gluon condensate, at the same time gives rise to a confining static fermionic potential, following the  criterion for quark confinement of 't Hooft \cite{tH03, tH003, tH07}. Finally, in the conclusions, we discuss some possible future paths to this research.

\section{BRST transformations}
We start by defining a complex gauge field $\mathcal{A}_{\mu }$ and its associated ghost $c$ transforming in the usual way under BRST
\begin{eqnarray}
s\mathcal {A}_{\mu }&=&-(\partial_{\mu }c-ig[\mathcal{A}_{\mu },c]) ,\nonumber \\
sc &=& -igc^{2}; \label{sA}
\end{eqnarray}
and the complex conjugated gauge field $\bar \mathcal{A}_{\mu }$ and its associated ghost $\bar{c}$ transforming as 

\begin{eqnarray}
s\bar \mathcal {A}_{\mu }&=&-(\partial_{\mu }\bar{c}-ig[\bar \mathcal {A}_{\mu },\bar{c}]) , \nonumber \\
s\bar{c} &=& -ig\bar{c}^{2} . \label{sAbar}
\end{eqnarray}

We can also define the usual curvature $\mathcal{F}_{\mu \nu}$ associated to $\mathcal{A}_{\mu }$
\begin{eqnarray}
\mathcal{F}_{\mu \nu}(\mathcal{A})=\partial_\mu \mathcal{A}_\nu - \partial_{\nu } \mathcal{A}_\mu -ig[\mathcal{A}_\mu, \mathcal{A}_\nu],
\label{F}
\end{eqnarray}
and its complex conjugate $\bar \mathcal{F}_{\mu \nu}$
\begin{eqnarray}
\bar \mathcal{F}_{\mu \nu}(\bar \mathcal{A})=\partial_\mu \bar \mathcal{A}_\nu - \partial_{\nu } \bar \mathcal{A}_\mu -ig[\bar \mathcal{A}_\mu, \bar \mathcal{A}_\nu].
\label{Fbar}
\end{eqnarray}
With these definitions, as anticipated in the Introduction, the element $ Tr \mathcal{F} \bar {\mathcal{F}} $ is not invariant under the BRST transformations. Only holomorphic elements as $ Tr \mathcal{F}^2$ and $ Tr {\bar \mathcal{F}}^2$ will be invariant under the set (\ref{sA}, \ref{sAbar}).


Up to this point, we have been following the standard definition of a complex gauge field theory. Now we come to the transformations of a pair of real scalar fields, $\varphi$ and $\psi$ in the adjoint representation, where we use the freedom that the complex theory allows us,
\begin{eqnarray}
 s\varphi &=& ig\varphi c -ig\bar{c}\varphi  , \nonumber \\
 s\psi &=& ig\psi \bar{c}-ig c\psi  . \label{spsi}
\end{eqnarray}
It is easy to confirm that the BRST operator $s$ is still nilpotent with these definitions, and that the object $ Tr\psi\varphi$ becomes invariant under its action, which will be useful to the construction of our potential and, ultimately, of the condensate. These transformations (\ref{spsi}) can be rewritten in terms of commutators and anticommutators as follows

\begin{eqnarray}
 s\varphi &=& i {g \over 2}\{\varphi , c - \bar{c}\}+i{g \over 2} [ \varphi , c + \bar{c} ] , \nonumber \\
 s\psi &=& i{g \over 2} \{ \psi , \bar{c} - c \} +i{g \over 2} [ \psi , \bar{c} + c ]   . \label{spsi2}
\end{eqnarray}
In this form, it becomes transparent the presence of the anticommutators in the scalar field transformations. This makes clear the problem that we pointed out after eqs.(\ref{phigaugetransf},\ref{phibargaugetransf}). As anticommutators are not in general closed operations inside the vector space of a Lie algebra, we must define how (\ref{spsi2}) makes sense. Planning to work with a complex $sl(N,C)$ algebra, we know that its fundamental representation matrices expand the most general  $N \times N$ matrix space together with the $N \times N$ identity. Then, transformations (\ref{spsi2}), and consequently (\ref{spsi}), make sense if we admit that the scalar field $\varphi$, together with the $\varphi^{i}$, $i=1,...,N^2 -1$, components associated to each $sl(N,C)$ generator, also have a $\varphi^{0}$ component associated to the $N \times N$ identity (and the same for $\psi$). 

Once we established the field content of the theory, and the BRST structure, we can proceed with the construction of the dynamics. The next step is to define the coupling among scalar and gauge fields, i.e., we need covariant derivatives. They are

\begin{eqnarray}
  \nabla_{\mu } \varphi &=& \partial_{\mu }\varphi + ig\varphi \mathcal{A}_{\mu }-ig \bar{\mathcal{A}}_{\mu }\varphi ,\nonumber \\
  {\nabla_{\mu } }\psi &=& {\partial_{\mu }\psi}+ig \psi \bar{\mathcal{A}}_{\mu }-ig \mathcal{A}_{\mu }\psi , \label{nablapsi}
\end{eqnarray}
which transform as

\begin{eqnarray}
  s\nabla_{\mu } \varphi &=& ig(\nabla_{\mu } \varphi ) c -ig \bar{c}(\nabla_{\mu } \varphi )\nonumber \\
  s{\nabla_{\mu } }\psi&=&ig({\nabla_{\mu } \psi})\bar{c}-ig c({\nabla_{\mu } \varphi}), \label{snablapsi}
\end{eqnarray}
following the covariance established by (\ref{spsi}). It is convenient to show the covariant derivatives explicitly in terms of commutators and anti-commutators

\begin{eqnarray}
 \nabla_{\mu } \varphi &=& \partial_{\mu }\varphi +\frac{ig}{2} ( -\{ \bar{\mathcal{A}}_{\mu }- \mathcal{A}_{\mu },\varphi \} - [\bar{\mathcal{A}}_{\mu }+ \mathcal{A}_{\mu },\varphi ] )\nonumber \\
 \nabla_{\mu } \psi &=& \partial_{\mu }\psi +\frac{ig}{2} (\{ \bar{\mathcal{A}}_{\mu }- \mathcal{A}_{\mu },\psi \}-[\bar{\mathcal{A}}_{\mu }+ \mathcal{A}_{\mu },\psi ] ). \label{comnablapsi}
\end{eqnarray}
These expressions (\ref{comnablapsi}) will be useful in the description of the masses for the vector fields after the choice of the non-symmetric vacua, a mechanism that we build in the following section.

\section{Invariant action and i-particles}

Gathering the elements that we have been discussing so far, we write the BRST invariant action

\begin{eqnarray}
 S= \int d^{4}x (\frac{i}{4}\mathcal{F}^{a}_{\mu\nu}\mathcal{F}^{a}_{\mu\nu}-\frac{i}{4}\bar{\mathcal{F}}^{a}_{\mu\nu}\bar{\mathcal{F}}^{a}_{\mu\nu} + Tr(\nabla_{\mu}\varphi)(\nabla_{\mu}\psi)  + V(\varphi,\psi )  )+ S_{GF} ,
 \label{action}
\end{eqnarray}
with $\mathcal{F}$ and ${\bar{\mathcal{F}}}$ the curvatures defined in (\ref{F}) and (\ref{Fbar}) respectively.
The gauge fixing sector $S_{GF}$ will soon have our attention, but first we present the scalar fields potential with the non-symmetric vacua,

\begin{eqnarray}
 V(\varphi, \psi )= -\frac{m^{2}}{2} \varphi^{a} \psi^{a}+\frac{\lambda }{4}(\varphi^{a} \psi^{a})^{2}   .
 \label{potential}
\end{eqnarray}
This potential has minima along the condition
\begin{equation}
  <\varphi^{a} \psi^{a}>= \frac{m^{2}}{\lambda }.
  \label{minima}
\end{equation}
Once  a specific vacuum is chosen, the scalar fields will develop vacuum expectation values (vevs)
\begin{eqnarray}
\varphi&\mapsto& \varphi + \mu \nonumber \\
\psi&\mapsto& \psi + \mu  \nonumber \\
Tr(\mu^{2}) &=&\frac{m^{2}}{2\lambda}
\label{vev}
\end{eqnarray}
Notice that, for simplicity, we considered equal vevs for both fields, which is not mandatory. With this choice, we can expand the interaction sector among scalar and gauge fields of the action (\ref{action}) around this vacuum, obtaining the characteristic bilinear terms of a symmetry breaking process 
\begin{eqnarray}
  (\frac{ig}{2} )^{2} Tr(-\{ \bar{\mathcal{A}}_{\mu }- \mathcal{A}_{\mu },\mu \}\{ \bar{\mathcal{A}}_{\mu }- \mathcal{A}_{\mu },\mu \} + [\bar{\mathcal{A}}_{\mu }+ \mathcal{A}_{\mu },\mu ][\bar{\mathcal{A}}_{\mu }+ \mathcal{A}_{\mu },\mu ] ),
  \label{Amass}
\end{eqnarray}
where eqs.(\ref{comnablapsi}) were used. These terms will give massive poles to the gluons propagators, but the specific nature of the resultant physics will obviously depend on the original gauge group and on the specific form of the $\mu$ vacuum.  From now on, we will focus on the example of a $SL(3,C)$ gauge invariance. This means that the scalar field $\varphi=\varphi^{a}T^{a}$ (and analogously $\psi$) now have components from $a=0$ to $a=8$, with $T^{i}$,  $i=1,...,8$, representing the eight Gell-Mann matrices, and $T^{0}=\sqrt{ \frac{1}{6}} I$, with $I$ the 3x3 identity matrix. In this way, the usual relation $tr(T^{a}T^{b})= \frac {1}{2} \delta^{ab}$ is preserved. The gauge fields $A_{\mu }$ and $\bar{A}_{\mu }$ then have eight complex components each, projected also on the eight $T^{i}$ Gell-Mann matrices. Notice that the presence of anticommutators in (\ref{Amass}) will bring consequences rather different than the usual for the gluon poles. For instance, as the anticommutators are not tensors of the gauge algebra, we highlight that the masses that the eight complex gauge fields $A_{\mu }^{a}$ (and of their conjugate $\bar{A}_{\mu }^{a}$) will develop after the choice of $\mu$ are deeply dependent of this selection of Gell-Mann matrices in the construction of the fields.

Before specifying the $\mu$ vacuum, we display the gauge fixing $S_{GF}$. It is well known that in a symmetry breaking process, the adequate gauge fixing is the 't Hooft gauge. It is a fundamental tool in order to study the physical content of the broken phase while explicitly retaining the renormalizability properties of the theory. We will adapt this gauge fixing to our complex case. In the end, as we will see, it will be extremely useful in the calculation of the condensate. Then, to implement this gauge fixing we begin by introducing two gauge conditions
\begin{eqnarray}
G&=& \partial_{\mu}\mathcal{A}_{\mu} +\frac{g\alpha}2\left(\psi\mu-\mu\varphi-\frac13Tr\left\{\psi\mu-\mu\varphi\right\} I \right)\nonumber \\
\bar G&=& \partial_{\mu}\bar{\mathcal{A}}_{\mu}+\frac{g\alpha}2 \left(\mu\psi-\varphi \mu-\frac 13Tr\left\{\mu\psi-\varphi \mu\right\} I \right),
\label{Gcondition}
\end{eqnarray}
where $\alpha$ is a gauge parameter,  and a pair of anti-ghosts $q$ and $\bar{q}$, and their respective Lagrange multipliers $b$ and $\bar b$, transforming in BRST doublets

\begin{eqnarray}
 s q= -ib, &&  s b=0, \nonumber \\
 s\bar{q}= i\bar b, &&  s \bar b=0.
 \label{antighosts}
\end{eqnarray}

Our gauge fixing takes the form

\begin{equation}
S_{GF}= s\int d^{4}x \left( Tr\left(-2q{G} -2\bar q\bar G+\alpha  qb +\alpha \bar{q}\bar b\right)\right),
\label{GF}
\end{equation}
so that, once we integrate out the Lagrange multipliers $b$ and $\bar b$, we get from (\ref{GF}), besides Faddeev-Popov like terms, the contribution  $\frac i{\alpha}Tr \left( G^2   -{\bar G}^2\right)$ to the action. Then, using the definitions in (\ref{Gcondition}), we see that this element will cancel the term mixing the gauge fields and the Goldstone bosons in the action (\ref{action}) after a vev $\mu$ is chosen, one of the main goals of the 't Hooft gauge. Also, we will be left with the standard gauge fixing   $\frac{i}{2\alpha}( (\partial_{\mu}{\mathcal{A}}^{a}_{\mu})^{2}-(\partial_{\mu}\bar{\mathcal{A}}^{a}_{\mu})^{2})$, appropriated to the definition of the gauge fields propagators.

After completing the presentation of our gauge fixed invariant action, we can now discuss the effect of the background $\mu$ vacuum. There are some inequivalent choices of this vacuum, representing different phases of this theory, some of them leading to the generation of i-particles. In what follows, we select a specific phase where we will show the presence of these fundamental objects. But actually, this phase stands out among the others mainly for the existence of a condensate with a K\"all\'{e}n-Lehmann like propagation, which will be the subject of the next section. This vacuum is given by

\begin{equation}
\mu = \frac{2 \nu }{\sqrt{3}} (\sqrt{2}T^{8} - T^{0}), \; \; \; \; \; \;   \nu = \sqrt{\frac{m^{2}}{4 \lambda}},
\label{vacuum}
\end{equation}
satisfying (\ref{vev}). In this phase, the gauge fields  $\mathcal{A}_{\mu }^{a}$ and $\bar{\mathcal{A}}_{\mu }^{a}$, with $a=(1,2,3)$, will remain massless, reflecting a SL(2,C) residual invariance of the vacuum. In fact, as can be easily seen by direct inspection, these directions commute and anti-commute with that of the vacuum (\ref{vacuum}),  which is the sense of gauge invariance in this theory. In the directions $a=(4,5,6,7)$ we will find the i-particles, with propagators

\begin{eqnarray}
<\bar{\mathcal{A}}_{\mu}^{a}\bar{\mathcal{A}}_{\nu}^{a}> &=&   \frac{i}{k^{2}+ig^{2}\nu^{2}}\theta_{\mu\nu} +\frac{i\alpha}{k^{2}+i\alpha g^{2}\nu^{2}}\omega_{\mu\nu} ,\nonumber \\
<\mathcal{A}_{\mu}^{a}\mathcal{A}_{\nu}^{a}> &=& - \frac{i}{k^{2}-ig^{2}\nu^{2}}\theta_{\mu\nu} -\frac{i\alpha}{k^{2}-i\alpha g^{2}\nu^{2}}\omega_{\mu\nu} ,\nonumber \\
\label{prop4567}
\end{eqnarray}
where
\begin{eqnarray}
\theta_{\mu\nu} &=& \delta_{\mu\nu}-\frac{k_{\mu}k_{\nu}}{k^{2}},\nonumber \\
\omega_{\mu\nu} &=& \frac{k_{\mu}k_{\nu}}{k^{2}} .
\end{eqnarray}
These i-particles are not particles in the standard sense, as they do not have asymptotic physical states. This is still a consequence of the loss of positivity of our starting action, i.e., of its complex gauge symmetry. But as explained in the Introduction, such i-particles were identified as the building blocks of a Hilbert space in theories based on the development of Gribov's ideas for confinement. Their emergence here is most welcome.

The last direction,  $a=8$,  commutes with the vaccum (\ref{vacuum}), but obviously does not have a trivial anti-commutator with it. This is the peculiarity of this theory, and we see that the propagators along this direction have a different form from the previous,

\begin{eqnarray}
 <\bar{\mathcal{A}}^{8}_{\mu}\bar{\mathcal{A}}^{8}_{\nu}> &=& \frac{ik^{2}+ \frac{4}{3} g^{2}\nu^{2}}{k^{4}} \theta_{\mu\nu}+ \frac{i\alpha k^{2}+ \frac{4}{3}\alpha^{2} g^{2}\nu^{2}}{k^{4}} \omega_{\mu\nu} , \nonumber \\
<\bar{\mathcal{A}}^{8}_{\mu} \mathcal{A}^{8}_{\nu}> &=& \frac{4 g^{2}\nu^{2}}{3k^{4}}\{\theta_{\mu\nu} +\alpha^{2}\omega_{\mu\nu} \}, \nonumber \\
<\mathcal{A}^{8}_{\mu} \mathcal{A}^{8}_{\nu}> &=& -\frac{ik^{2}- \frac{4}{3} g^{2}\nu^{2}}{k^{4}}\theta_{\mu\nu}- \frac{i\alpha k^{2}- \frac{4}{3}\alpha^{2} g^{2}\nu^{2}}{k^{4}}\omega_{\mu\nu} .
\end{eqnarray}
These propagators, although inappropriate for a gluon condensation, have a well studied form, playing a special role in the Wilson loop approach of fermion confinement. The simultaneous  presence of both kind of gauge propagators in the same phase is the novelty of this theory. We will explore this property when we couple this theory to fermions in section V. For now, we will continue with this scalar-vector field sector, exploring a possible condensate. In the next sections we will take the gauge parameter as $\alpha =1$ .

 \section{A gluon condensate}

The absence of positive definite elements in the action (\ref{action}) is an expected consequence of a theory based on a complex gauge group. The new gauge transformation for the scalar fields as proposed in (\ref{spsi}) allowed the concurrence of ${\mathcal{A}}$ and $\bar{\mathcal{A}}$ in some of the terms of the scalar-gauge interaction, as we found in (\ref{Amass}). Anyway, this still does not assure the positiveness of any invariant object at the level of the action. The search for a physical positive norm space must be done in composite objects, as condensates, possibly with an apparent higher dimension. The idea is that once the vev $\mu$ is fixed, the dimension of such object must get limited to that of spacetime. Then we hope to find a candidate  mixing together scalar fields, attaining their background vevs, together with gauge fields with the i-particle structure just met. For this kind of mixing, we can get inspiration in 't Hooft-Polyakov's abelian projection of a scalar-gauge (Georgi-Glashow) theory \cite{tH74,Pol74}, a mechanism responsible for quark confinement in three dimensions as shown in \cite{Pol77}. The whole object should be gauge invariant, and observing the gauge transformations defined by (\ref{sA}), (\ref{sAbar}) and (\ref{spsi}), we can find such a generalization as follows

\begin{eqnarray}
O(x)= Tr(\varphi \mathcal{F} \psi \bar{\mathcal{F}}) . \label{ox}
\end{eqnarray}
In the phase defined by the vacuum of (\ref{vacuum}), this operator becomes

\begin{eqnarray}
 O(x) &=&{2\nu^2\over 3}\mathcal{F}^8_{\mu\nu} \bar{\mathcal{F}}^8_{\mu\nu} .
\label{o}
\end{eqnarray}
We find that in the phase of (\ref{vacuum}), the operator $O(x)$ becomes positive definite. Now we can study the two-point correlation function $  \langle O(k) O(-k) \rangle $ of this operator. We intend to show that this correlator presents a K\"allén-Lehmann spectral representation, with a positive spectral density    
\begin{equation}
  \langle O(k) O(-k) \rangle =\int_{\tau_{0} }^{\infty }d\tau \frac{\rho(\tau )}{\tau +k^{2}}, \label{oKL}
\end{equation}
with  
\begin{equation}
  \rho(\tau ) \geq 0     \,  ,  \,  \tau_{0} \leq \tau \leq  \infty . 
\end{equation}
As we are assuming that in this phase the theory enters a confinement regime, we should look for such spectral representation in the dominant contribution in the gauge coupling $g$. This is given by the $g^{4}$ element
 
\begin{eqnarray} 
 \langle O(k) O(-k) \rangle _{g^{4}} &=& \frac{\nu^4 g^4}{4} \int \frac{d^4p}{(2\pi)^4}\int \frac{d^4q}{(2\pi)^4}\int \frac{d^4l}{(2\pi)^4}I^{o}(q,l,+ig^{2}\nu^{2} )I^{0}(p,\varsigma , -ig^{2}\nu^{2} )\nonumber \\
 I^{0}(q,l,+ig^{2}\nu^{2} )&=& \frac{1}{(q^{2}+ig^{2}\nu^{2})(l^{2}+ig^{2}\nu^{2}) }\nonumber \\
 I^{0}(p,\varsigma , -ig^{2}\nu^{2} )&=&\frac{1}{(p^{2}-ig^{2}\nu^{2})(\varsigma ^{2}-ig^{2}\nu^{2}) }\nonumber \\
 \varsigma &=&k-p-q-l \;. \label{ogc2}
 \end{eqnarray}
 
In order to obtain the spectral density in the K\"all\'{e}n-Lehmann representation it is just enough to rewrite the integral in a form where it becomes apparent. We can use the known result regarding the correlator of a pair of i-particles (see \cite{Bau10} for this and other details of the following calculations)
\begin{equation}
\int \frac{d^4l}{(2\pi)^4}\frac{1}{(l^{2}+ig^{2}\nu^{2})((k-p-q-l)^{2}-ig^{2}\nu^{2}) }  =\int_{t_{0} }^{\infty }dt \frac{\rho(t )}{t +(k-p-q)^{2}},
\end{equation} 
where

\begin{eqnarray}
t_{0}&=&2g^{2}\nu^{2} ,\nonumber \\
\rho({t})&=&\frac{1}{16 \pi^2}\frac{\sqrt{t^{2}-4g^{4}\nu^{4}}}{t},
\end{eqnarray}
so that in (\ref{ogc2}) we obtain
\begin{eqnarray}
\langle O(k) O(-k) \rangle _{g^{4}}&=&\int_{t_{0} }^{\infty }dt \rho(t ) \int \frac{d^4p}{(2\pi)^4}\frac{1}{t +(k-p)^{2}} C(p),\nonumber \\
C(p)&=&\int \frac{d^4q}{(2\pi)^4}\frac{1}{(q^{2}+ig^{2}\nu^{2})((p-q)^{2}-ig^{2}\nu^{2}) }.
\end{eqnarray}
Again, using the same procedure for $C(p)$
\begin{equation}
\langle O(k) O(-k) \rangle _{g^{4}} = \int_{t_{0} }^{\infty }dt \rho(t ) \int \frac{d^4p}{(2\pi)^4}\frac{1}{t +(k-p)^{2}}\int_{t _{0} }^{\infty }dv \frac{\rho(v )}{v +p^{2}} .
\end{equation}
Finally,
\begin{equation}
\langle O(k) O(-k) \rangle _{g^{4}} = \int_{t_{0} }^{\infty }dt \rho(t )\int_{t _{0} }^{\infty }dv \rho(v ) \int_{(\sqrt{t}+\sqrt{v})^{2} }^{\infty }dr \frac{\rho_{1}(r,t,v)}{r+k^{2}} ,
\end{equation}
in which the branch cut for two real masses needs the integration from $(\sqrt{t}+\sqrt{v})^{2} $ and \cite{Dud11}

\begin{equation}
\rho_{1}(r,t,v) = \frac{1}{16 \pi^2}\frac{\sqrt{[r - (v + t)]^{2}-4v t }}{r},
\end{equation}

Now we need to perform the convolution of spectral representations. In order to do that we will first take into account the last two integrals
\begin{equation}
G (k)=\int_{t _{0} }^{\infty }dv \rho(v ) \int_{(\sqrt{t}+\sqrt{v})^{2} }^{\infty }dr \frac{\rho_{1}(r,t,v)}{r+k^{2}},   
\end{equation}
where $\sqrt{t}\geq  \sqrt{t_{0}}$. Performing the substitution $\sqrt{\omega } =\sqrt{t } + \sqrt{v}$ we obtain that 
\begin{equation}
G (k)= \int_{(\sqrt{t_{0}}+\sqrt{t})^{2} }^{\infty } d\omega \bar{\rho }(\omega ) \int_{\omega }^{\infty }dr\frac{\rho_{1}(r,t,\omega)}{r+k^{2}}. \label{G}
\end{equation}
The new spectral density $\bar{\rho }(\omega )$ can be written in terms of the old one as
\begin{equation}
\bar{\rho }(\omega )=\frac{\sqrt{w}-\sqrt{t }}{\sqrt{w}} \rho [(\sqrt{w}-\sqrt{t })^{2} ] ,
\end{equation}
and its positivity  $ \bar{\rho }(\omega )\geq 0 $ is still assured when $ \omega \geq  (\sqrt{t_{0}}+\sqrt{t })^{2} $. Now we can change the order of integration in (\ref{G})
\begin{equation}
G (k)= \int_{(\sqrt{t_{0}}+\sqrt{t})^{2} }^{\infty } dr\frac{1}{r+k^{2}} \int_{(\sqrt{t_{0}}+\sqrt{t})^{2} }^{\infty }{d\omega \bar{\rho }(\omega )\rho_{1}(r,t,\omega)} ,
\end{equation}
arriving at the form

\begin{equation}
 G (k)=\int_{(\sqrt{t_{0}}+\sqrt{t })^{2}}^{\infty } dr \frac{\rho_{2}(r,t)}{r+k^{2}} ,
\end{equation}
where
\begin{equation}
\rho_{2}(r,t) = \int_{(\sqrt{t_{0}}+\sqrt{t})^{2} }^{\infty }{d\omega \bar{\rho }(\omega )\rho_{1}(r,t,\omega)} \geq 0.
\end{equation}
Inserting this result again on $ \langle O(k) O(-k) \rangle _{g^{4}} $ and performing the same procedure we obtain
\begin{equation}
  \langle O(k) O(-k) \rangle _{g^{4}} = \int_{4t_{0} }^{\infty }dr\frac{\rho_{3}(r)}{r+k^{2}}, \label{OO}
\end{equation}
with 

\begin{equation}
 \rho_{3}(r) = \int_{4t_{0} }^{\infty }{d\omega \bar{\rho }(\omega )\rho_{2}(r,\omega)}\geq 0  . \label{rho3}
\end{equation}
Equations (\ref{OO}) and (\ref{rho3}) state the desired K\"all\'{e}n-Lehmann form of the correlator (\ref{ogc2}).

\section{The fermionic potential}
We begin this section by describing the coupling of ${\mathcal{A}}$ and $\bar{\mathcal{A}}$  to a set of fermionic fields $\xi$, $\chi$ and their conjugated fields $\bar{\xi}$, $\bar{\chi}$ in the action 

\begin{eqnarray}
S_F&=e^{-i\theta}\bar{\chi}\left(\gamma^{\mu}D_{\mu} + iM\right)\xi + e^{i\theta}\bar{\xi}\left(\gamma^{\mu}\bar{D}_{\mu} + iM\right)\chi , \label{SF}
\end{eqnarray}
where the covariant derivatives are
\begin{eqnarray}
	(D_{\mu}\xi)^i= \partial_{\mu}\xi^i-ig(T^a)_{ij}\mathcal{A}_{\mu}^a\xi^j \, ,
	\hspace{1cm}  (\bar{D}_{\mu}\chi)^i= \partial_{\mu}\chi^i-ig(T^a)_{ij}\bar{\mathcal{A}}_{\mu}^a\chi^j \, . \label{DF}
\end{eqnarray}
The action (\ref{SF}) is invariant under the BRST transformations (\ref{sA}), (\ref{sAbar}) and these for the fermionic fields

\begin{eqnarray}
	s\xi^i&=&-ig(T^a)_{ij}c^a\xi^j \, ,  \hspace{1cm} s\bar{\chi}^i=ig\bar{\chi}^j(T^a)_{ji}{c}^a \, ,\nonumber \\
	s\chi^i&=&-ig(T^a)_{ij}\bar{c}^a\chi^j  \, ,
	\hspace{1cm} s\bar{\xi}^i=ig\bar{\xi}^j(T^a)_{ji}\bar{c}^a \, .
\end{eqnarray}
In this sense the derivatives (\ref{DF}) are covariant
\begin{eqnarray}
	 s(D_{\mu}\xi)^i=-ig(T^a)_{ij}c^a(D_{\mu}\xi)^j \, ,
	\hspace{1cm}  s(\bar{D}_{\mu}\chi)^i=-ig(T^a)_{ij}\bar{c}^a(\bar{D}_{\mu}\chi)^j   \, .
\end{eqnarray}
At this point we must call attention to the $\theta$ coefficients appearing in front of the elements in the action (\ref{SF}). In principle, they are allowed by the fact that each element of (\ref{SF}) is independently gauge invariant, the reality of the action being the only request to relate them. In the end, as we will see, they will play an interesting physical role in the fermionic potential. 

The method for the computation of the interquark potential in the one-gluon-exchange approximation is described in  \cite{Lu91,Lu96}. In \cite{At19} this method was applied to the case of the GZ theory, showing the non-confining character of the GZ gluon propagator, just confirming a result that was already known for some time by other means \cite{Gra09}. The potential is obtained as a Fourier transformation

\begin{eqnarray}
V(\vec{x})=-(2\pi)^3\int d^3k e^{-i\vec{k}\cdot \vec{x}}T_{fi}(k), \label{V}
\end{eqnarray}
of  the scattering amplitude $T_{fi}$ defined in terms of the $S$-matrix as

\begin{eqnarray}
S_{fi}=<f,out\mid i,in>=\delta_{fi}+i(2\pi)^4\delta^4(P_f-P_i)T_{fi}.
\end{eqnarray}

The standard procedure is to consider the heavy-quark (non-relativistic) approximation, and that the quarks inside a hadron are in a colour-singlet state \cite{Lu91}. In our case, we identify four different graphic contributions to the one-gluon exchange scattering amplitude (noticing that, as usual, all pair annihilation contributions to this one-gluon approximation vanish), i.e.,

\begin{eqnarray}
T_{fi}= {1\over{(2\pi)^6}} (t_1+t_2+t_3+t_4). \label{Tfi}
\end{eqnarray}

\begin{figure}[ht]\label{fig1}
	\centering
	\begin{minipage}[b]{0.25\linewidth}
		\centering
		\includegraphics[width=0.7\textwidth]{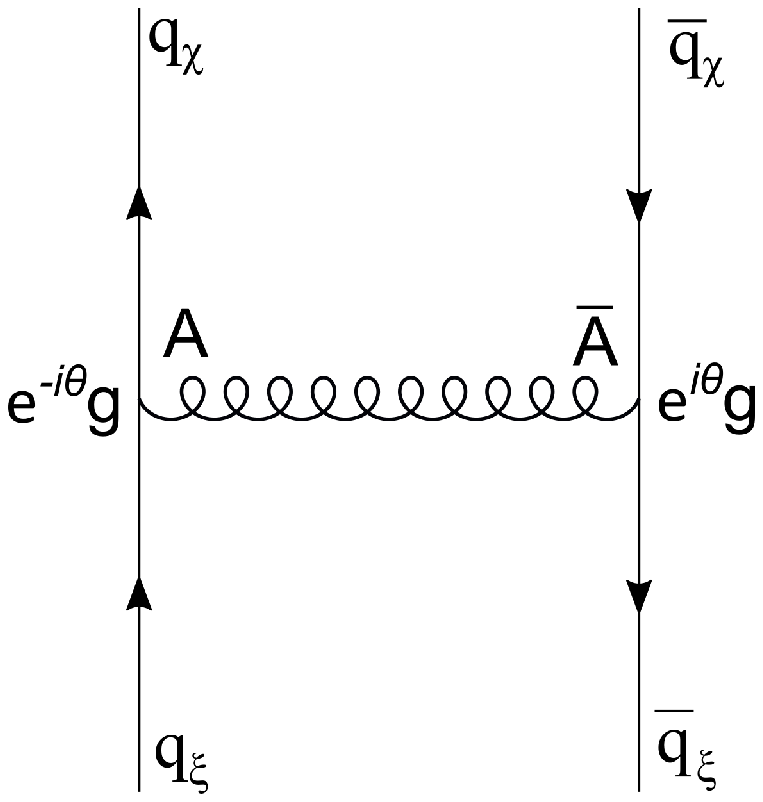}
		\caption{$t_1$-Incoming $\xi$ particle and $\xi$ antiparticle}
		\label{fig1}
	\end{minipage}
	\hspace{0.5cm}
	\begin{minipage}[b]{0.25\linewidth}
		\centering
		\includegraphics[width=0.7\textwidth]{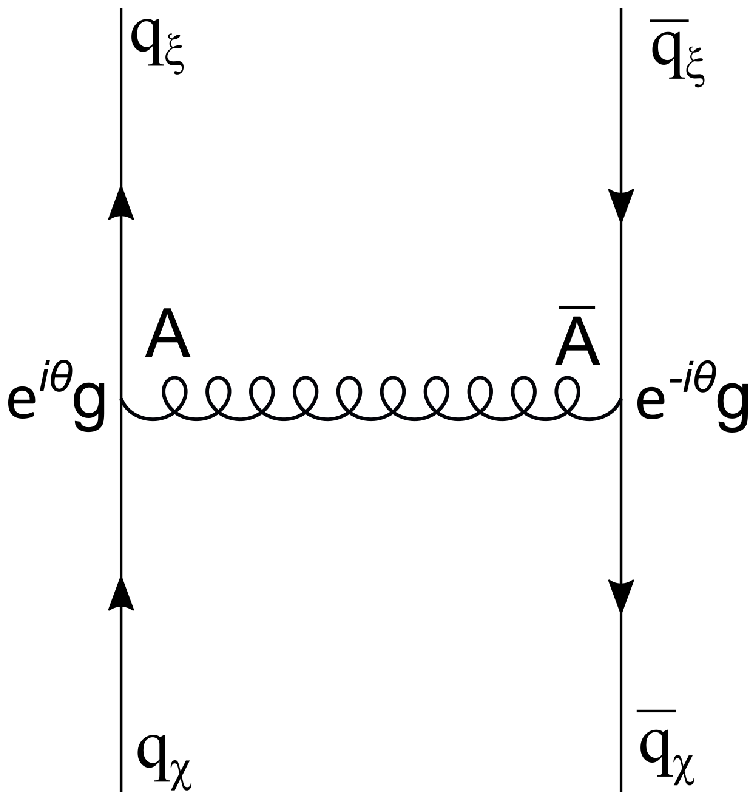}
		\caption{$t_2$-Incoming $\chi$ particle and $\chi$ antiparticle}
		\label{fig2}
	\end{minipage}
\end{figure}

\begin{figure}[ht]\label{fig1}
	\centering
	\begin{minipage}[b]{0.25\linewidth}
		\centering
		\includegraphics[width=0.7\textwidth]{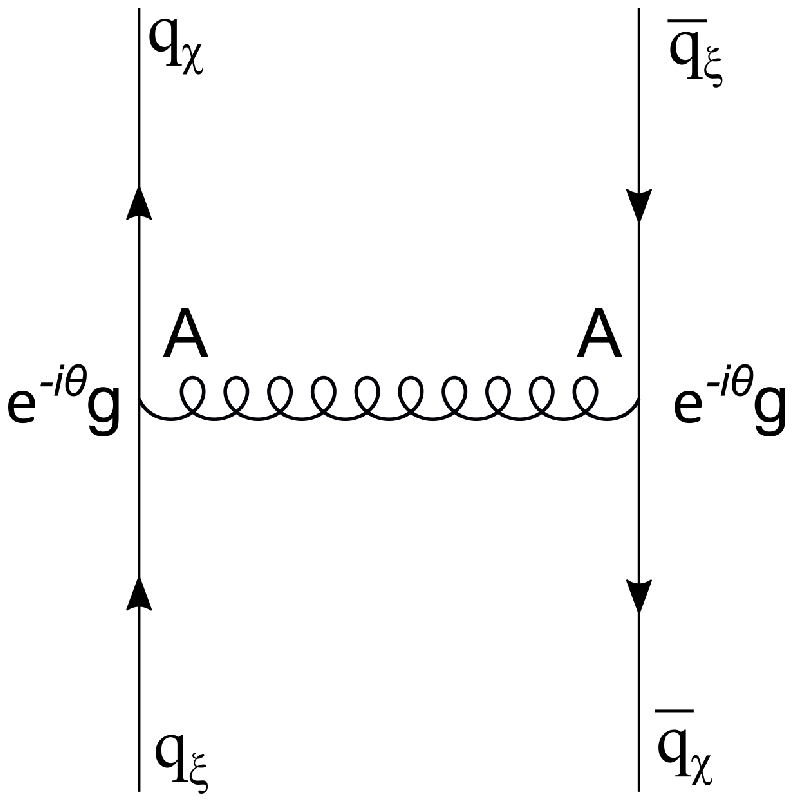}
		\caption{$t_3$-Incoming $\xi$ particle and $\chi$ antiparticle}
		\label{fig3}
	\end{minipage}
	\hspace{0.5cm}
	\begin{minipage}[b]{0.25\linewidth}
		\centering
		\includegraphics[width=0.7\textwidth]{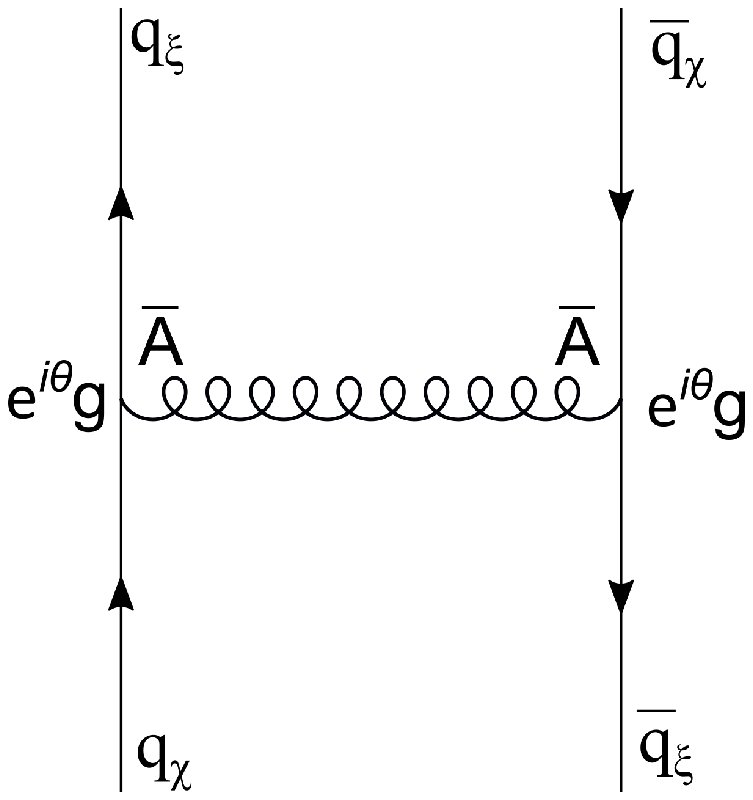}
		\caption{$t_4$-Incoming $\chi$ particle and $\xi$ antiparticle}
		\label{fig4}
	\end{minipage}
\end{figure}

The contributions from $t_1$ and $t_2$ give the same result

\begin{eqnarray}
t_1=t_2=  {\frac{2g^4{{\nu}^2}}{9}}{1\over{k^4}},
\end{eqnarray}
whereas the last two contributions are the complex conjugate of each other

\begin{eqnarray}
t_3 = t_4^\ast ={\frac{2g^2e^{-2i\theta}}{3}}\left(-{{i}\over{k^2}}-{{i}\over{k^2-ig^2v^2}} +{} {{g^2v^2}\over{3k^4} }  \right) .
\end{eqnarray}

Then, the scattering amplitude (\ref{Tfi}) is

\begin{eqnarray}
T_{fi}=\frac{2g^2}{3(2\pi)^6}\left[\frac{g^2{v^2}}{3k^4}+e^{-2i\theta}\left(-{{i}\over{k^2}}-{{i}\over{k^2-ig^2v^2}} + {{g^2v^2}\over{3k^4} }  \right)+c.c.\right] \, ,
\end{eqnarray}
which, using (\ref{V}),  leads to the interquark potential

\begin{equation}
V(\vec{x})=\frac{g^2}{3\pi}\left[x\frac{g^2v^2\left(1+\cos2\theta\right)}6+
\frac1x\left(	{\sin2\theta}+
e^{-\frac{\sqrt 2}{2} xgv} \sin\left({2\theta-\frac{\sqrt 2xgv}2}\right)\right)
	\right] \, . \label{Vconf}
\end{equation}

In this potential (\ref{Vconf}) we find the confining linear dependence, together with a Coulomb like contribution, which characterizes a potential of the Cornell type \cite{Ei78, Ei80, Ei02}. Then we understand that the role played by the  $\theta$ coefficients appearing in the action (\ref{SF}) is associated to spatially compact states, i.e., to the size of the quark-bags.

\section{Conclusions}

We have introduced here the study of a QCD based on a complex group, defining a complex gauge field theory. Our main intention was to show that, although not well suited for Yang-Mills theory in a perturbative regime, a gauge theory with a complex symmetry develops some of the features required for the description of a confined phase. In particular, such theory seems to be the natural environment for the generation of i-particles, fundamental constituents of propagators in the Gribov scenario.
In our present case, we explored the freedom that is present in the way that scalar fields may transform under the action of a complex group. After the choice of a non-symmetric vacuum, this theory leads to some gluons with propagations characteristic of i-particles, together with a gluon propagating with the mode predicted by 't Hooft for the quark confinement. With this, we have outlined in the same theory a possible gluon condensate, together with a confining quark potential after the coupling to fermions. This is a novelty, as, up to now, all theories of the GZ kind do not generate quark confinement.

Now we should stress that the picture presented here is just a sketch of the work that needs to be faced in order to build up a consistent theory. Several points must be addressed. For example, one should immerse this structure inside a larger theory in order to actually describe a symmetry breaking mechanism. In fact, up to now we have not said a word about the previous symmetric phase before the choice of the vacuum expectation value of the scalar fields, when the i-particles show up. But we can at least argument on a general characteristic of this symmetric phase. We imagine it to be described by a topological field theory. The reason for this is not only for the close relationship that complex gauge theory has with topology when it appears in the literature, as we pointed out in the Introduction (as can be seen for example in \cite{Wit91,Wit10}).  Mainly because once we observe that the action (\ref{action}) prior to the symmetry breaking already lacks positivity, we understand that the immersion that we have been suggesting for this phase must lead to the recovery of the energy bound. This can be achieved by the increase of the level of symmetry for this symmetric phase, possibly a high level of supersymmetry leading to a topological field theory. Actually, this is not altogether new, as a theory with some of these characteristics has already been described as the low energy field theory limit of M-theory in three dimensions \cite{BLMP}. There, a theory with a twisted Chern-Simons action reduces to the difference of two standard Chern-Simons terms with two gauge fields. Then, the accomplishment of these ideas would possibly allow the association of the observable condensates after symmetry breaking to topological invariants of the symmetric phase.
 
Also, only after this development one will be able to aim at the renormalization problem, for the condensate, potential, and for the whole theory. Other points are the search for new condensates and the use of different symmetry groups, which may be adequate to settle the discussion of a grand unified theory in this scenario.

As a final comment, we call attention to the residual SL(2,C) which remains as a symmetry of the vacuum (\ref{vacuum}). Its algebra is isomorphic to the Lorentz group algebra so(3,1), and may be expected to occur after the breaking of higher SL(N,C) groups. Anyway, this interesting feature here remains as a mere coincidence.


\begin{thebibliography}{10}


\bibitem{Gribov}V.~N.~Gribov, ``Quantization of Nonabelian Gauge Theories,''
  Nucl.\ Phys.\ B {\bf 139}, 1 (1978).


\bibitem{Zw89}D.~Zwanziger,
  ``Local and Renormalizable Action From the Gribov Horizon,''
  Nucl.\ Phys.\ B {\bf 323}, 513 (1989).
  
\bibitem{Zw12}N. Vandersickel and D.~Zwanziger,
  ``The Gribov problem and QCD dynamics,''
  Phys. \ Rept. {\bf 520} 175 (2012)
[arXiv:1202.1491 [hep-th]].

\bibitem{Dud08}D.~Dudal, J.~A.~Gracey, S.~P.~Sorella, N.~Vandersickel and H.~Verschelde,
``A Refinement of the Gribov-Zwanziger approach in the Landau gauge: Infrared propagators in harmony with the lattice results,''
 Phys.\ Rev.\ D {\bf 78}, 065047 (2008)
[arXiv:0806.4348 [hep-th]].

\bibitem{Zw93}D.~Zwanziger,
``Renormalizability of the critical limit of lattice gauge theory by BRS invariance,''
 Nucl.\ Phys.\ B {\bf 399}, 477 (1993).
 
\bibitem{So09} L. Baulieu and S. P. Sorella,
``Soft breaking of BRST invariance for introducing non-perturbative infrared effects in a local and
renormalizable way,''  Phys. Lett. B {\bf 671}, 423-496 (2009), [arXiv:0808.1356 [hepth]].


\bibitem{Fj83}K. Fujikawa,  
  ``Dynamical stability of the BRS supersymmetry and the Gribov problem,''
   Nuclear Physics B  {\bf 223} 218-234 (1983) . 

\bibitem{Ted11} M. A. L. Capri, A. J. Gomez, M. S. Guimaraes, V. E. R. Lemes,
S. P. Sorella and D. G. Tedesco, ``Renormalizability of the linearly broken
formulation of the BRST symmetry in presence of the Gribov horizon
in Landau gauge Euclidean Yang-Mills theories,''
Phys. Rev. D {\bf 83}, 105001 (2011)
[arXiv:1102.5695 [hep-th]].

\bibitem{Lav11}P.M. Lavrov, O.V. Radchenko and A.A. Reshetnyak,
  ``Is soft breaking of BRST symmetry consistent?,''
  JHEP {\bf 1110} 043 (2011)
  [arXiv:1108.4820 [hep-th]].

\bibitem{Lav12}P.M. Lavrov, O.V. Radchenko and A.A. Reshetnyak,  
  ``Soft breaking of BRST symmetry and gauge dependence,''
   Mod. \ Phys, \ Lett. \ A {\bf 27} 1250067 (2012) 
   [arXiv:1201.4720 [hep-th]].


\bibitem{Lav15}P.M. Lavrov and O.V. Radchenko,
  ``Once again on the Gribov horizon,''
  Mod. \ Phys. \ Lett. \ A {\bf 30} 1550215 (2015) 
  [arXiv:1509.03031 [hep-th]].
  
\bibitem{Sor09}S.~P.~Sorella,
  ``Gribov horizon and BRST symmetry: a few remarks,''
  Phys. Rev. D {\bf 80}  025013 (2009)
    [arXiv:0905.1010 [hep-th]].
    
    
\bibitem{Kon09}K. Kondo
   ``The nilpotent BRST symmetry for the Gribov-Zwanziger theory,''
   [arXiv:0905.1899 [hep-th]].   

\bibitem{Cap15}M. A. L. Capri, D. Dudal, D. Fiorentini, M. S. Guimaraes, I. F. Justo, A. D. Pereira, B. W. Mintz, L. F. Palhares,
R. F. Sobreiro and S. P. Sorella, 
 ``Exact nilpotent nonperturbative BRST symmetry for the Gribov-Zwanziger action in the
linear covariant gauge,'' 
 Phys. Rev., D {\bf 92} , no. 4, 045039 (2015)
  [arXiv:1506.06995 [hep-th]].


\bibitem{Cap16}M. A. L. Capri, D. Dudal, D. Fiorentini, M. S. Guimaraes, I. F. Justo, A. D. Pereira, B. W. Mintz, L. F. Palhares,
R. F. Sobreiro and S. P. Sorella, 
``A local and BRST-invariant Yang-Mills theory within the Gribov horizon,''
Phys. Rev. D {\bf 94}, no. 2, 025035 (2016) 
[arXiv:1605.02610 [hep-th]].

\bibitem{Del86} R. Delbourgo and G. Thompson,
``Massive, Unitary, Renormalizable Yang-Mills Theory without Higgs Bosons,''
 Phys. Rev. Lett. {\bf 57}, 2610 (1986).

\bibitem{Kubo87} J. Kubo,
``Comment on "Massive, Unitary, Renormalizable Yang-Mills Theory without Higgs Bosons",''
 Phys. Rev. Lett. {\bf 58}, 2000 (1987).


\bibitem{Kos87} P. Kosi\u{n}ski and L. Szymanowski,
``Comment on "Massive, Unitary, Renormalizable Yang-Mills Theory without Higgs Bosons",''
 Phys. Rev. Lett. {\bf 58}, 2001 (1987).
 
 
 \bibitem{Del88} R. Delbourgo, S. Twisk and G. Thompson,
``Massive Yang-Mills Theory:  Renormalizability versus Unitarity,''
Int. J. Mod. Phys. A {\bf 3} 435 (1988).


\bibitem{Rue04}H. Ruegg and M. Ruiz-Altaba,
``The Stueckelberg Field,''
Int. J. Mod. Phys. A {\bf 19}, 3265 (2004)
 [arXiv:0304245 [hep-th]].
 
\bibitem{Cap18} M. A. L. Capri, D. Dudal, M. S. Guimaraes, A. D. Pereira, B. W. Mintz, L. F. Palhares, and S. P. Sorella,
``The universal character of Zwanziger’s horizon function in Euclidean Yang-Mills theories,'' 
Phys. Lett., B {\bf 781}, 48 (2018)
[arXiv:1802.04582 [hep-th]].

   
\bibitem{Mag94}N. Maggiore and M. Schaden, 
   Landau gauge within the Gribov horizon, 
   Phys. Rev. D  {\bf 50} 6616–6625 (1994)  
    [arXiv:9310111 [hep-th]].
    
\bibitem{Vil11}L. C. Q. Vilar, O. S. Ventura and V. E. R. Lemes, 
Phys. Rev. D  {\bf 84} 105026 (2011) 
[arXiv:1108.3305 [hep-th]].
    
\bibitem{Cuc11}A. Cucchieri and T. Mendes, Parallel talk at The XXVIII International Symposium on Lattice Field Theory,
Lattice2010, June 14-19, 2010, Villasimius, Italy, 7 pp, [arXiv:1101.4537 [hep-lat]].
    
    
\bibitem{Sc15}M. Schaden and D.~Zwanziger,
   ``Living with spontaneously broken BRST symmetry. I. Physical states and cohomology,'' 
   Phys. Rev. D {\bf 92} 025001 (2015)
     [arXiv:1312.4823 [hep-th]].    
    
    
    

\bibitem{Sor12}D.~Dudal and S.~P.~Sorella,    
    ``The Gribov horizon and spontaneous BRST symmetry breaking,''
   Phys. Rev. D {\bf 86} 045005  (2012)
     [arXiv:1205.3934 [hep-th]].
     
     
  

     
\bibitem{Bau10}L. Baulieu, D. Dudal, M. S. Guimaraes, M. Q. Huber, S.~P.~Sorella,
N. Vandersickel and D. Zwanziger
 ``Gribov horizon and i-particles: about a toy model and the construction of physical operators,''
 Phys. Rev. D {\bf 82}, 025021 (2010)
 [arXiv:0912.5153 [hep-th]]. 
 
\bibitem{Cap11}M.~A.~L.~Capri, A.~J.~Gomez, M.~S.~Guimaraes, V.~E.~R.~Lemes, S.~P.~Sorella and D.~G.~Tedesco,
  ``Constructing local composite operators for glueball states from a confining Gribov propagator,''
  Eur.\ Phys.\ J.\ C {\bf 71}, 1525 (2011)
  [arXiv:1009.3062 [hep-th]].
  
\bibitem{Dud10} D. Dudal, M. S. Guimaraes and S. P. Sorella, 
    ``Glueball masses from an infrared moment problem and nonperturbative Landau gauge,''
   Phys. Rev. Lett. {\bf 106}, 062003  (2011)
   [arXiv:1010.3638 [hep-th]]. 
  
  
\bibitem{Dud11}D.~Dudal and M.~S.~Guimaraes,
  ``On the computation of the spectral density of two-point functions: complex masses, cut rules and beyond,''
  Phys.\ Rev.\ D {\bf 83}, 045013 (2011)
  [arXiv:1012.1440 [hep-th]].
 
 
\bibitem{Sor10}S.~P.~Sorella,
  ``Gluon confinement, i-particles and BRST soft breaking,''
  J.\ Phys.\ A {\bf 44}, 135403 (2011)
  [arXiv:1006.4500 [hep-th]].
  
 \bibitem{Wu76}T. T, Wu and C. N. Yang,
   ``Static Sourceless gauge field,'' 
  Phys.\ Rev.\ D {\bf 13}, 3233 (1976).
  
\bibitem{Wit91}E.~Witten,
  ``Quantization of Chern-Simons gauge theory with complex gauge group,''
   Comm. Math. Phys. {\bf 137}, 29 (1991).

\bibitem{Wit10}E.~Witten,
  ``Analytic continuation Of Chern-Simons theory,''
   in Chern-Simons Gauge Theory: 20 Years After, p. 347. AMS, (2010)   
  [arXiv:1001.2933 [hep-th]].
  
\bibitem{Bau09}L. Baulieu, 
  ``N = 4 Yang–Mills theory as a complexification of the N = 2 theory,'' 
  Nucl. Phys. Proc. Suppl. {\bf 192-193}, 27 (2009) 
  [arXiv:0906.1289 [hep-th]].
  
\bibitem{tH03} G. 't Hooft,
``Perturbative confinement,'' 
 Nucl. Phys. Proc. Suppl. {\bf 121}, 333 (2003)
 	[arXiv:0207179 [hep-th]].

 \bibitem{tH003} G. 't Hooft, 
``Confinement of quarks,'' 
 Nucl. Phys. A {\bf 721}, 3 (2003).
 
 \bibitem{tH07}G. 't Hooft, 
``Models for confinement,''  
 Prog. Theor. Phys. Suppl. {\bf 167}, 144 (2007).
 
 
 \bibitem{tH74} G. 't Hooft, ``Magnetic monopoles in unified gauge theories,'' 
 Nuc. Phys. B {\bf 79}  276, (1974).
  
 \bibitem{Pol74} A. M. Polyakov, ``Particle spectrum in quantum Field theory,'' 
 JETP Letters {\bf 20},  194 (1974).
 
 \bibitem{Pol77} A.M. Polyakov, ``Quark confinement and topology of gauge theories,''
  Nucl. Phys. B {\bf 120}, 429 (1977).
  
 \bibitem{Lu91} W. Lucha, F. F. Schoberl and D. Gromes, ``Bound states of quarks,''
  Phys. Rept. {\bf 200}, 127 (1991). 


\bibitem{Lu96}W. Lucha  and F. F. Schöberl, ``Effective potential models for hadrons,'' 
[arXiv:9601263 [hep-th]].


\bibitem{At19}A. Cucchieri, T. Mendes  and W. M. Serenone, ``Lattice gluon propagator and one-gluon-exchange potential,''  
Braz. J. Phys. {\bf 49}, 548 (2019)
[arXiv:1704.08288 [hep-lat]]. 

\bibitem{Gra09}J.A. Gracey, ``The one loop MSbar static potential in the Gribov-Zwanziger Lagrangian,''
JHEP {\bf 1002}, 009 (2010)
[arXiv:0909.3411 [hep-th]]. 


\bibitem{Ei78}E. Eichten, K. Gottfried, T. Kinoshita, K. Lane  and T.M.
Yan, ``Charmonium: the model,'' Phys. Rev. D {\bf 17}, 3090 (1978), {\bf 21}, 313 (1980) (erratum).

\bibitem{Ei80}E. Eichten, K. Gottfried, T. Kinoshita, K. Lane and T.M.
Yan, ``Charmonium: Comparison with Experiment,'' Phys. Rev. D {\bf 21}, 203 (1980).

\bibitem{Ei02}E.J. Eichten, K. Lane and C. Quigg, ``B-Meson gateways to missing charmonium levels,'' Phys. Rev. Lett. {\bf 89},
162002 (2002) [arXiv:0206018 [hep-ph]].

\bibitem{BLMP}J. Bagger, N. Lambert, S. Mukhi, C. Papageorgakis, ``Multiple Membranes in M-theory,''
Phys. Rept. {\bf 527},  1-100 (2013) [arXiv:1203.3546 [hep-th]].



 




\end{thebibliography}
\end{document}